\documentclass[prl,aps,superscriptaddress,twocolumn,floatfix,nofootinbib]{revtex4-2}

\usepackage{mathtools}
\usepackage[usenames,dvipsnames]{xcolor}
\usepackage{amsmath}
\usepackage{amssymb,mathrsfs}
\usepackage{graphicx}
\usepackage[colorlinks=true]{hyperref}

\newcommand{\prt}{\partial}

\newcommand{\om}{\omega}

\newcommand{\ox}{\overline{x}}

\begin{document}

\title{Propagation of instability fronts in modulationally unstable systems}

\author{A. M. Kamchatnov}  \affiliation{Institute of Spectroscopy,
  Russian Academy of Sciences, Troitsk, Moscow, 108840, Russia}

\author{D. V. Shaykin}  
  \affiliation{Moscow Institute of
  Physics and Technology, Institutsky lane 9, Dolgoprudny, Moscow
  region, 141700, Russia}

\begin{abstract}
We study evolution of pulses propagating through focusing nonlinear media.
Small disturbance on the smooth initial non-uniform background leads to
formation of the region of strong nonlinear oscillations. We develop here
an asymptotic method for finding the law of motion of the front of this
region. The method is applied to the focusing nonlinear Schr\"{o}dinger
equation for the particular cases of Talanov and
Akhmanov-Sukhorukov-Khokhlov initial distributions with zero initial phase. 
The approximate analytical results agree well with the exact numerical 
solutions for these two problems.
\end{abstract}

\pacs{05.45.−a, 42.60.Jf, 42.65.Sf}


\maketitle

{\it 1. Introduction.} Modulational instability of initially smooth wide wave beams was
first observed \cite{pr-65} in the form of filamentation of light beams propagating
through nonlinear media. This phenomenon was explained in Ref.~\cite{bt-66} with the
use of the focusing nonlinear Schr\"{o}dinger (NLS) equation
\begin{equation}\label{eq1}
  i\psi_t+\frac12\psi_{xx}+|\psi|^2\psi=0
\end{equation}
written here in standard non-dimensional variables and in 1D geometry, $\psi$ being
the wave field variable. Bespalov and Talanov obtained the dispersion law
\begin{equation}\label{eq2}
  \om(k)=k\sqrt{k^2/4-\rho}
\end{equation}
for linear waves $\propto\exp[i(kx-\om t)]$ propagating along a uniform background
$\psi=\psi_0=\sqrt{\rho}$ and noticed that the frequency becomes complex for
$0<k<2\sqrt{\rho}$. This means instability of small perturbations of such a background,
and the growth of these unstable modes leads eventually at the nonlinear stage of
evolution to filamentation of the beam.

Discovered independently \cite{bf-67} modulational instability of Stokes water waves
drew much attention and it was followed by intense study of different aspects of this
phenomenon (see, e.g., Ref.~\cite{zo-09} for the early history of these studies).
In particular, Whitham and Lighthill \cite{lighthill-65} noticed that if one takes
into account a nonlinear correction $\om(k)\mapsto\om(k)+\om_1(k)a^2$ to the dispersion law,
where $a$ is the wave amplitude, then the group velocity of wave packets splits and its two
values become complex if the signs of $\om_1(k)$ and of the group velocity dispersion
$\om^{\prime\prime}(k)$ are opposite. Thus, the condition for modulational instability
of envelopes of wave packets in the long wavelength limit reads
\begin{equation}\label{eq3}
  \om_1(k)\om^{\prime\prime}(k)<0,
\end{equation}
and this condition fulfills for deep enough water waves in agreement with observations of
Benjamin and Feir~\cite{bf-67} which demonstrated instability of envelopes of Stokes waves.

These remarkable results posed the problem of evolution of initially smooth wave packets (or
wide light beams) in a focusing nonlinear medium. For example, numerical experiments (see, e.g.,
\cite{mk-98,bk-99,ct-02}) showed that an initially smooth localized distribution focuses forming
a spike at some moment of time, and after that a region of large amplitude nonlinear
oscillations appears at the center of the distribution and this region spreads with time
outward over the smooth evolving background distribution. Formation of the region of large 
nonlinear oscillations
can be stimulated by a localized disturbance which violates smoothness of the initial
distribution \cite{kk-68,karpman-75}. As was noticed in Ref.~\cite{lighthill-65}, if the
initial distributions are represented
by analytical functions, then evolution before the focusing time can be described by an
analytical theory similar to standard methods of compressible fluid dynamics. However,
a general enough description of the region of nonlinear oscillations is a much more
difficult task. In particular cases of completely integrable evolution equations like the
NLS equation (\ref{eq1}), the powerful inverse scattering transform method \cite{zs-71}
can be used for derivation of some important characteristics of this region.
In particular, a quasi-classical approximation was developed for its description, see,
e.g., Refs.~\cite{kmlm-03} and references therein. Being mathematically rigorous,
this method does not provide simple enough formulas for motion of the instability front
that could be used in practice and a need for a simpler approach is obvious. Our aim in this
Letter is to present such an approach to finding an asymptotic form of the law of motion
of the instability front applicable to both completely integrable and non-integrable equations.

{\it 2. General theory.}
We assume that the system under consideration is described by two wave
variables $\rho(x,t)$ and $u(x,t)$ which we call the `density' and `flow velocity',
correspondingly. As was shown in Ref.~\cite{sg-69}, in many physical situations the
equations of wave dynamics can be cast to the form
\begin{equation}\label{eq4}
   \rho_t+(\rho u)_x=0,\quad
   (\rho u)_t+(\rho u^2+P)_x=0,
\end{equation}
where $P=P(\rho,u,\rho_x,\rho_t,u_x,u_t,\rho_{xx},\ldots)$ and higher order derivatives
in $P$ correspond to dispersive (or dissipative) effects. In dispersionless limit,
these higher order derivatives are disregarded, $P=P(\rho,u)$, and Eqs.~(\ref{eq4})
take the form of equations of compressible fluid dynamics. The system is modulationally unstable
in the long wavelength limit if $\prt P(\rho,u)/\prt\rho<0$. For example, the
substitution $\psi=\sqrt{\rho}\exp\left(i\int^xu(x',t)dx'\right)$ transforms
the NLS equation (\ref{eq1}) into the system
\begin{equation}\label{eq5}
  \begin{split}
  & \rho_t+(\rho u)_x=0,\\
  & u_t+uu_x-\rho_x+\left(\frac{\rho_x^2}{8\rho^2}-\frac{\rho_{xx}}{4\rho}\right)_x=0,
  \end{split}
\end{equation}
and in dispersionless limit $P=-\rho^2/2$, $dP/d\rho=-\rho<0$.
Such dispersionless limiting equations describe the
self-focusing of initial distributions in the geometric optics approximation.

Although, strictly speaking, in this case the geometric optics problem is ill-posed,
nevertheless, if the initial data are represented by analytic functions, then
solutions of dispersionless equations provide a very good approximation (see, e.g.,
\cite{mk-98,kmlm-03,miller-01}) up to the focusing moment and, as was mentioned above, 
they often can be
obtained by the classical methods of compressible fluid dynamics (see, e.g.,
\cite{lighthill-65,gs-70}). We assume that such a dispersionless solution is known and 
it is convenient for what follows to write this solution in an implicit form
\begin{equation}\label{eq6}
  x=x(\rho,u),\qquad t=t(\rho,u),
\end{equation}
so that solving this system with respect to $\rho$ and $u$ yields the time-dependent
distributions $\rho=\rho(x,t)$ and $u=u(x,t)$. This smooth dispersionless solution
becomes singular at the point $x_b$ at the focusing moment of time $t_b$ after which
this approximation breaks down since the derivatives $\rho_x(x,t_b),u_x(x,t_b)$ tend
to infinity as $x\to x_b$. Therefore, to avoid complications caused by inaccuracy of
the geometric optics approximation in the limit $t\to t_b$, we shall assume that the
instability is triggered by a small perturbation located in vicinity of the point
$x=0$ in the initial distribution. Then for $t>0$ the region of strong nonlinear
oscillations develops and, following the
ideas of the Gurevich-Pitaevskii theory of dispersive shock waves \cite{gp-1973}
(see also \cite{kamch-21b} and references therein), we assume that at the asymptotic
stage of evolution, when typical wavelength of these nonlinear waves is much smaller
than the length of the whole oscillatory region, the wave dynamics outside this region
can be described by the considered above dispersionless limit of the equations under
consideration with disregarded dispersion (diffraction) effects. Near the edges of the
oscillatory region the amplitude of oscillations is small and these waves propagate
along the smooth background (\ref{eq6}) according to the linearized Eqs.~(\ref{eq4})
which yield some dispersion law
\begin{equation}\label{eq7}
  \om=\om(k,\rho,u),
\end{equation}
where the wavelength  $2\pi/k$ is assumed much smaller than the characteristic size of
the background flow distributions. Then the edges of the oscillatory region
propagate with the group velocity
\begin{equation}\label{eq8}
  v_g=\frac{\prt\om}{\prt k}
\end{equation}
at some value of the wave number $k$. If we find this value of $k$, then we can
find the law of motion of the edges, that is of propagation of the instability fronts.

To determine the critical value of the wave number $k$ at the edges of the
oscillatory region, we shall consider first an example of the NLS equation case.
We notice that the
dispersion law  (\ref{eq2}) for $k>2\sqrt{\rho}$ has an inflection point at
$k_m=\sqrt{6\rho}$ at which the group velocity
\begin{equation}\label{eq9}
  v_g=\frac{d\om}{dk}=\frac{k^2-2\rho}{\sqrt{k^2-4\rho}}
\end{equation}
has its minimal value
\begin{equation}\label{eq10}
  v_m=2\sqrt{2\rho}.
\end{equation}
This value of the group velocity has an important physical meaning.
If we consider evolution of a small localized at $x=0$ disturbance
of the uniform distribution with constant $\psi_0=\sqrt{\rho}$, then we find
that the region of strong nonlinear oscillations appears around the origin
\cite{kk-68,karpman-75}. For not too large time of evolution, the region of oscillations
can be represented as a modulated periodic solution of Eq.~(\ref{eq1}) and the
evolution of the modulation parameters obeys the Whitham modulation equations
\cite{whitham-65,whitham-74} which were derived for the NLS equation in
Refs.~\cite{fl-86,pavlov-87}. Their appropriate self-similar solution was obtained
in Refs.~\cite{egkk-93,bk-94} and a similar problem for the one-dimensional
Heisenberg ferromagnet was solved in Ref.~\cite{kamch-92}. According to this theory,
the small-amplitude edges of the oscillatory region propagate with the group
velocity (\ref{eq10}). Recently this result was confirmed by a more rigorous
Riemann-Hilbert method in Refs.~\cite{bm-16,bm-17}.

The same situation holds for non-integrable
systems. For example, two-dimensional dark solitons of the defocusing NLS
equation are unstable with respect to bending perturbations (`snake' instability),
but they can be stabilized by a proper background flow. The condition for transition
to such stable states is that the flow velocity must be greater than the velocity
(\ref{eq5}) of the instability front propagation \cite{kp-08,kk-11}. Similar rule
for selection of front propagation velocity into unstable states was suggested long ago
\cite{dl-83} in the theory of patterns formation in dissipative systems and it was
confirmed in a number of various physical situations (see, e.g., \cite{saarloos-03}
and references therein).

Thus, we generalize these observations and assume that the front $x=x(t)$ of the oscillatory
region propagates into the unstable smooth region with the group velocity
\begin{equation}\label{eq11}
  \frac{dx}{dt}=v_m(\rho,u)
\end{equation}
corresponding to the inflection point of the dispersion law of linear waves
propagating along this smooth background. Naturally, the use of notion of the group velocity
means that our theory is asymptotic and is applicable only to the stage of evolution
later enough after the moment of formation of the region of nonlinear oscillations.

To put these ideas into practice, we assume for simplicity that the variable $u$ can
be excluded from the system (\ref{eq6}) and the function $v_m=v_m(\rho,u)$, so we get
\begin{equation}\label{eq12}
  x=x(\rho,t),\qquad v_m=v_m(\rho,t)
\end{equation}
where $\rho$ plays the role of the parameter along the path of the small amplitude
wave packet. Substitution of these expressions into Eq.~(\ref{eq11})
gives the differential equation
\begin{equation}\label{eq12b}
  \frac{d\rho}{dt}=\frac{v_m-\prt x/\prt t}{\prt x/\prt\rho}
\end{equation}
for variation the background density along the packet's path. In the case of instability
triggered by a small initial disturbance localized in vicinity of the point $x=0$,
this equation should be solved with the initial condition
\begin{equation}\label{eq13}
  \rho=\rho_0(0)\qquad\text{at}\qquad t=0,
\end{equation}
where $\rho=\rho_0(x)$ is the initial distribution of the density.
The resulting solution $\rho=\rho(t)$ of Eq.~(\ref{eq12b}) after substitution into 
Eq.~(\ref{eq12}) yields the asymptotic expression for the path of the edge.

Let us illustrate this approach by two examples when the evolution
of the field variable $\psi$ obeys the NLS equation (\ref{eq1}).

{\it 3. Talanov's self-focusing solution.} We take the initial distribution in the
form of a parabolic hump
\begin{equation}\label{eq15b}
 \rho_0(x)=
  \begin{cases}
    a^2(1-x^2/l^2), & \text{   \ $|x|\le l$}, \\
     0,       & \text{   \ $|x|>l$},
  \end{cases}
\end{equation}
and the initial phase is equal to zero everywhere, hence $u_0(x)=0$. The system
(\ref{eq5}) reduces in the dispersionless limit to the `inverted' shallow water
equations
\begin{equation}\label{70-2g}
   \rho_t+(\rho u)_x= 0,\qquad
     u_t+uu_x -\rho_x=0.
\end{equation}
As was shown by V.~Talanov \cite{talanov-65}, this system with the initial conditions
(\ref{eq15b}) has an exact solution in the form
\begin{equation}\label{eqc.147}
\begin{split}
  &\rho_b(x,t)=\frac{a^2}{f(t)}\left(1-\frac{x^2}{l^2f^2(t)}\right),\\
  &u_b(x,z)=\frac{f'(t)}{f(t)}x
  =-\frac{2a}{l}\frac{\sqrt{1-f}}{f^{3/2}}x,
  \end{split}
\end{equation}
where the function $f(t)$ is defined implicitly by the expression (see Supplemental Material
for details)
\begin{equation}\label{eqc.150}
  (2a/l)t=\sqrt{f(1-f)}+\arccos\sqrt{f}.
\end{equation}
This solution determines profiles of the density and the flow velocity as functions of time
up to the focusing moment $t_f=\pi l/(4a)$.

Now we suppose that the initial distribution (\ref{eq15b}) is perturbed and for
definiteness we take this disturbance in the form of a tiny hillock over the
density distribution at $x=0$. (In fact, the asymptotic evolution of the instability
region does not depend on the form of the disturbance.) As a result of such a disturbance,
the region of strong oscillations is formed, and our aim is to find the low of motion
of its edges.

The edge wave packet propagates upstream or downstream the background flow (\ref{eqc.147}) 
that has the local flow velocity $u_b(x,t)$, so the Doppler-shifted local dispersion law reads
\begin{equation}\label{eq19}
  \om(k)=k(u_b\pm\sqrt{k^2/4-\rho_b}),
\end{equation}
where the sign depends on the direction of propagation of the packet. As before, the
function $\om=\om(k)$ has an inflection point at $k=\pm\sqrt{6\rho_b}$ and the
corresponding group velocity equal to
\begin{equation}\label{eq20}
  v_g=u_b\pm2\sqrt{2\rho_b}.
\end{equation}
For definiteness, we shall consider the left instability front at $x<0$ for which
$u_b>0$ and $v_g<0$, that is we get with account of Eqs.~(\ref{eqc.147})
\begin{equation}\label{eq21b}
  \frac{dx}{dt}=-\frac{2a}{l}\frac{\sqrt{1-f}}{f^{3/2}}-
  2\sqrt{2}a\cdot\sqrt{\frac1f\left(1-\frac{x^2}{l^2f^2}\right)}.
\end{equation}
Formula (\ref{eqc.150}) is obtained by integration of the equation
\begin{equation}\label{eq22}
  \frac{df}{dt}=-\frac{2a}{l}\sqrt{\frac{1-f}{f}},
\end{equation}
so Eq.~(\ref{eq21b}) transforms to
\begin{equation}\label{eq23c}
  \frac{dx}{df}=\frac{x}f+\sqrt{2}l\cdot\sqrt{\frac{1-x^2/(lf)^2}{1-f}}.
\end{equation}
Due to the initial disturbance, the instability front starts its propagation at the point
$x=0$ at the moment $t=0$, when $f=1$, so Eq.~(\ref{eq23c}) should be solved with
the initial condition $x(f=1)=0$ and the solution reads
\begin{equation}\label{eq24b}
  x(f)=-lf\sin\left(\sqrt{2}\ln\frac{1+\sqrt{1-f}}{1-\sqrt{1-f}}\right).
\end{equation}
Together with Eq.~(\ref{eqc.150}), this equation determines the path of the
instability front in a parametric form $(x(f),t(f))$ with $f$ playing the role of
the parameter. The front's coordinate $x(f)$ touches the edge of the background
distribution (\ref{eqc.147}) when the argument of the sine-function in Eq.~(\ref{eq24b})
equals to $\pi/2$, that is at $f=f_c=1/\cosh^2(\pi/4\sqrt{2})\approx0.745$. The
background density vanishes at this point and our approach based on the assumption
that the small amplitude wave packet propagates along a nonzero background breaks
down. Thus, the parameter $f$ in Eq.~(\ref{eq24b}) changes in the interval $1\geq f>f_c$.

\begin{figure}[t]
\begin{center}
\begin{picture}(120.,100.)
\put(-55,0){\includegraphics[width = 4cm]{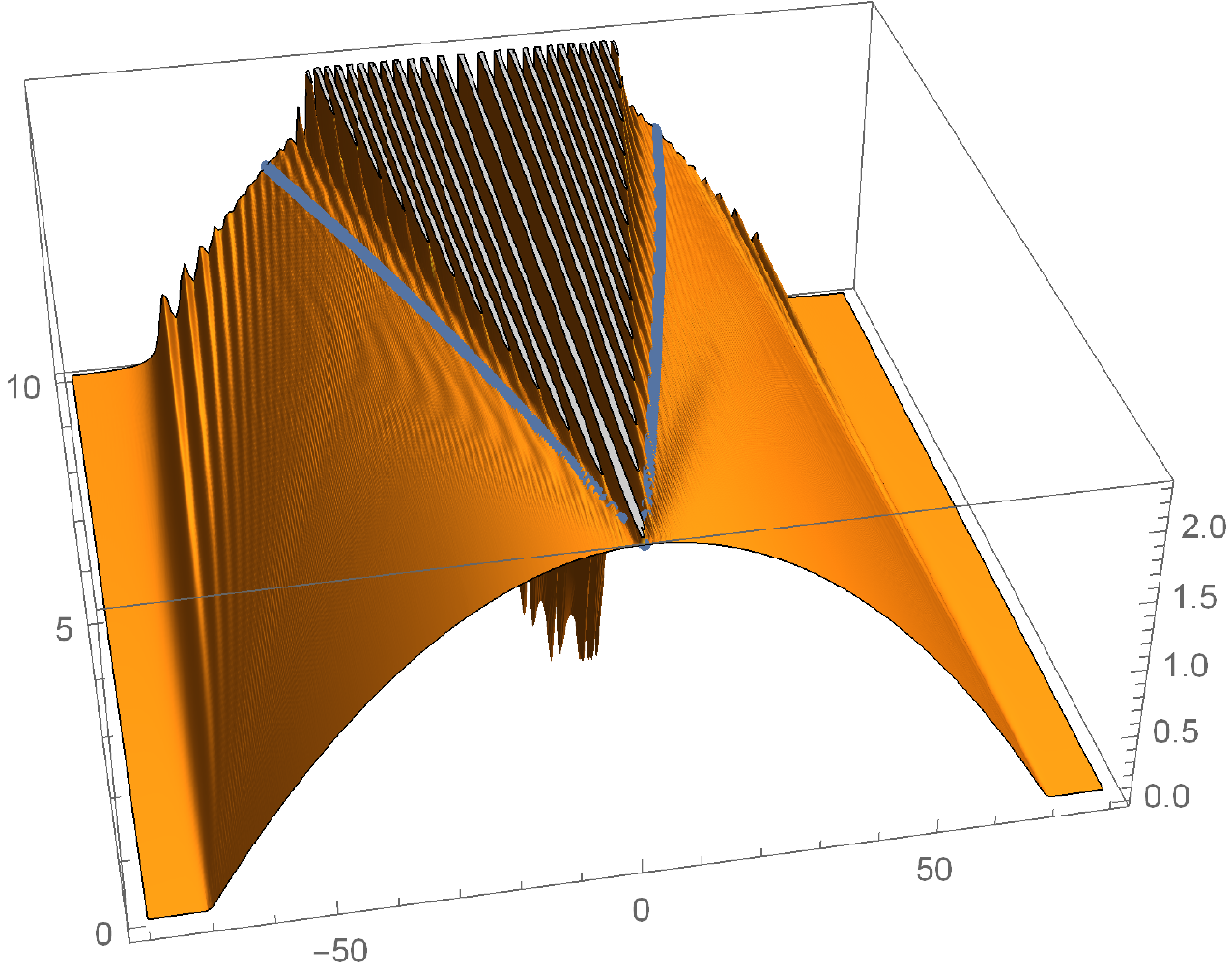}}
\put(65,0){\includegraphics[width = 4cm]{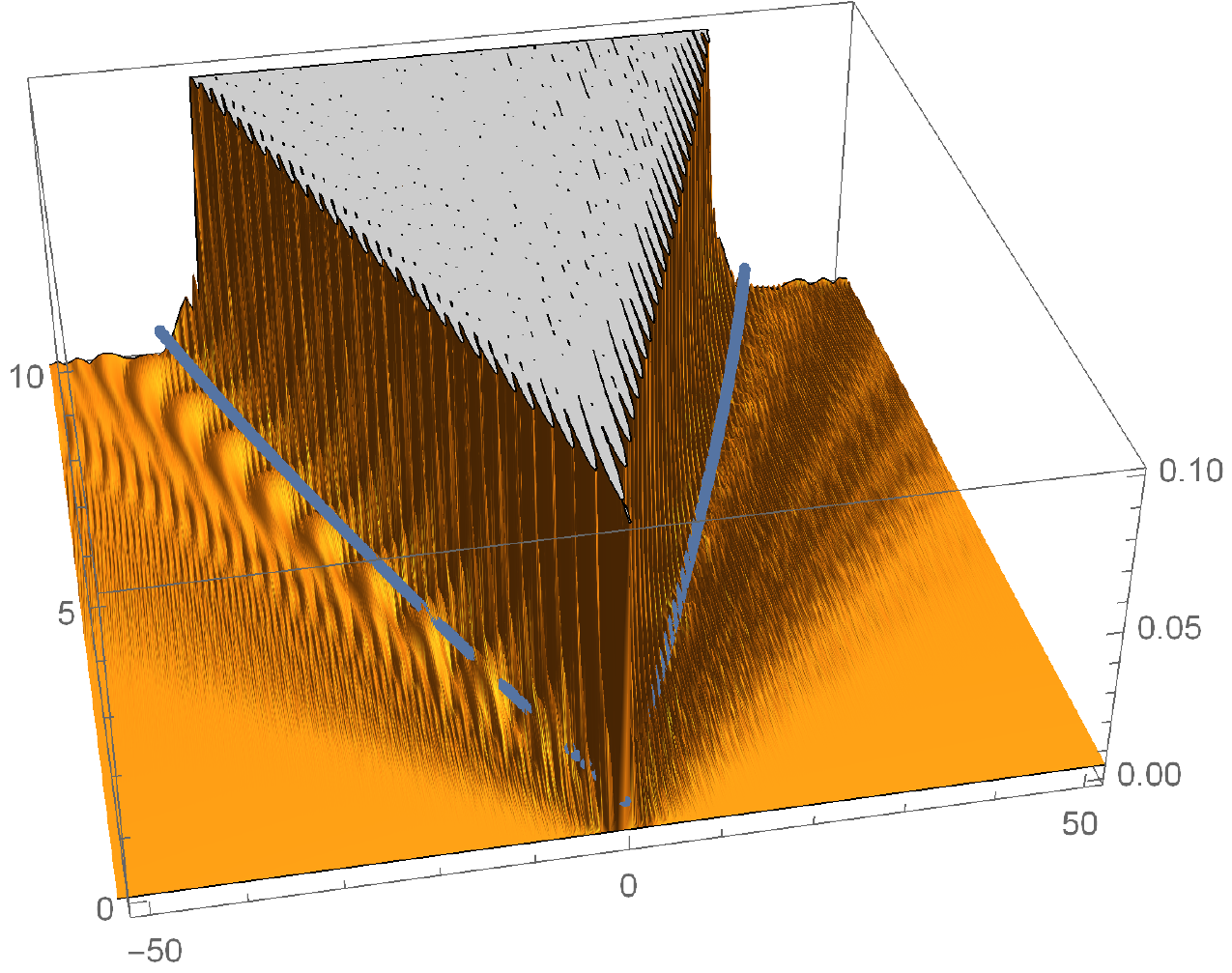}}
\put(35,80){(a)}
\put(155,80){(b)}
\end{picture}
\caption{(a) Plot of the numerical solution of the NLS equation (\ref{eq1}) for the
initial condition (\ref{eq15b}) with $a=1.5$, $l=70$, and a tiny perturbation
of density at $x=0$. (b) Plot of $|\rho(x,t)-\rho_b(x,t)|$ with subtracted background
evolution (\ref{eqc.147})
from the exact solution $\rho(x,t)$. Analytical paths of the instability fronts are
shown by blue lines.
 }
\label{fig1}
\end{center}
\end{figure}

We compare our analytical theory with the exact numerical solution of the NLS equation
with the disturbed parabolic initial distribution (\ref{eq15b}). In Fig.~\ref{fig1}(a) the
surface plot of the function $\rho(x,t)$ is shown where for clarity the large amplitude
peaks are cut at the level $\rho=2.3$. Theoretical paths of the instability fronts given
parametrically by the formulas $(\pm x(f),t(f),\rho_b(\pm x(f),t(f)))$ are shown by
blue lines. To show evolution of the region of large amplitude oscillations only,
we subtracted the background distribution $\rho_b(x,t)$ given by Eq.~(\ref{eqc.147}) from the
exact numerical solution $\rho(x,t)$ of the NLS equation and the corresponding surface
plot of $|\rho(x,t)-\rho_b(x,t)|$ is shown in Fig.~\ref{fig1}(b). Again the large amplitude
oscillations are cut for clarity at the level $|\rho(x,t)-\rho_b(x,t)|=0.1$ and the
fronts' paths are shown by blue lines. The theory does not contain any fitting
parameters and its agreement with the numerical solution seems quite good.

{\it 4. Akhmanov-Sukhorukov-Khokhlov self-focusing solution.} As another example,
let us consider the background evolution corresponding to the initial conditions
\begin{equation}\label{eq25b}
  \rho_0(x)=\frac{a^2}{\cosh^2(x/l)},\quad u_0(x)=0\quad\text{at}\quad t=0.
\end{equation}
Solution of this problem was obtained in Ref.~\cite{ask-66} and it can be written in the
following convenient for us parametric form (see Supplemental Material for details)
\begin{equation}\label{eq26}
  \rho(\xi,\eta)=a^2(1+\xi^2)(1-\eta^2),\quad u(\xi,\eta)=2a\xi\eta,
\end{equation}
\begin{equation}\label{eq27}
\begin{split}
  & t=\frac{l}a\cdot\frac{\xi}{(1+\xi^2)(1-\eta^2)},\\
  & x=l\left[\frac{2\xi^2\eta}{(1+\xi^2)(1-\eta^2)}-\frac12\ln\frac{1+\eta}{1-\eta}\right]
  \end{split}
\end{equation}
where $\xi$ and $\eta$ are parameters $(\xi\geq0, -1\leq\eta\leq1)$. Their elimination
yields the original implicit form of the solution found in Ref.~\cite{ask-66}:
\begin{equation}\label{eq28}
  \rho=\frac{a^2+\rho^2t^2/l^2}{\cosh^2[(x-ut)/l]},\quad u=-\frac{2\rho t}l\tanh\frac{x-ut}l.
\end{equation}
For us it is more convenient to exclude only the parameter $\eta$ to obtain
\begin{equation}\label{eq29b}
  \rho(\xi,t)=al\cdot\frac{\xi}t,\quad u(\xi,t)=2a\xi\sqrt{1-\frac{l\xi}{at(1+\xi^2)}}
\end{equation}
and
\begin{equation}\label{eq30b}
  x(\xi,t)=2at\xi\sqrt{1-\frac{l\xi}{at(1+\xi^2)}}-
  \frac{l}2\ln\frac{1+\sqrt{1-\frac{l\xi}{at(1+\xi^2)}}}
  {1-\sqrt{1-\frac{l\xi}{at(1+\xi^2)}}}.
\end{equation}
If we fix the moment of time $t$, then these formulas give us parametric forms of
distributions of the density $(\rho(\xi,t),x(\xi,t))$ and flow velocity $(u(\xi,t),x(\xi,t))$
as functions of $x$. This solution becomes singular in the limit $t\to t_b=l/(2a)$
with space derivatives of the distributions $\rho(x,t),u(x,t)$ tending to infinity.
Hence, the dispersive effects are essential in vicinity of this point so that singularity
is replaced here by a spike followed by formation of the region of oscillations.
Such a behavior of the solution of the NLS equation
was studied in Refs.~\cite{tvz-04,bgk-09,bt-13}. For $t>t_b$ not too close
to the singularity point the dispersionless solution (\ref{eq29b}), (\ref{eq30b}) remains
accurate enough outside the region of nonlinear oscillations and one can use it for
finding the paths of the instability fronts. However, dispersion effects introduce some
uncertainty in definition of the initial moment $t_b$ that is essential from a
practical point of view. To avoid these complications, we shall consider formation of the 
instability region due to small disturbance in the initial condition (\ref{eq25b}).

\begin{figure}[t]
\begin{center}
\begin{picture}(120.,100.)
\put(-55,0){\includegraphics[width = 4cm]{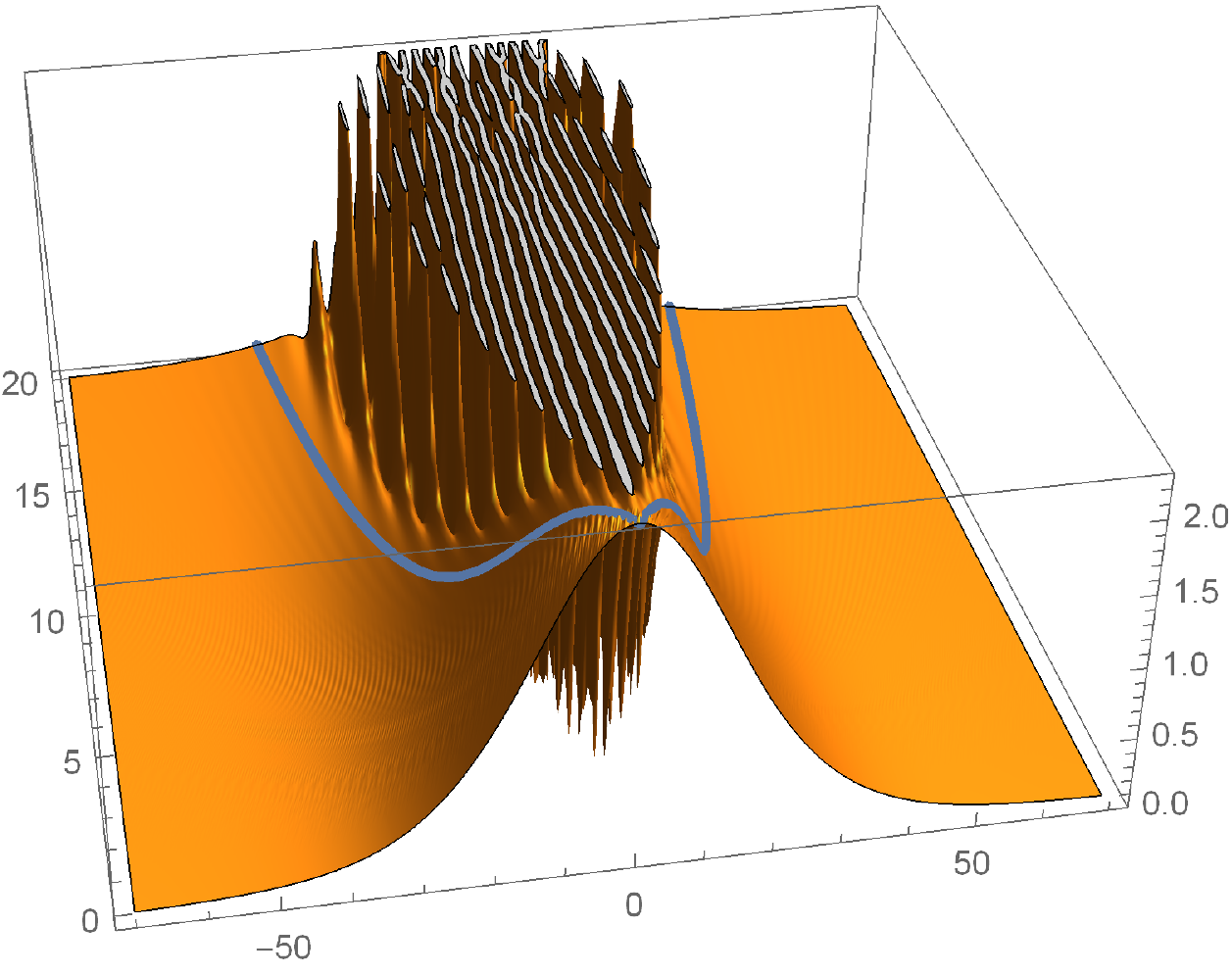}}
\put(65,0){\includegraphics[width = 4cm]{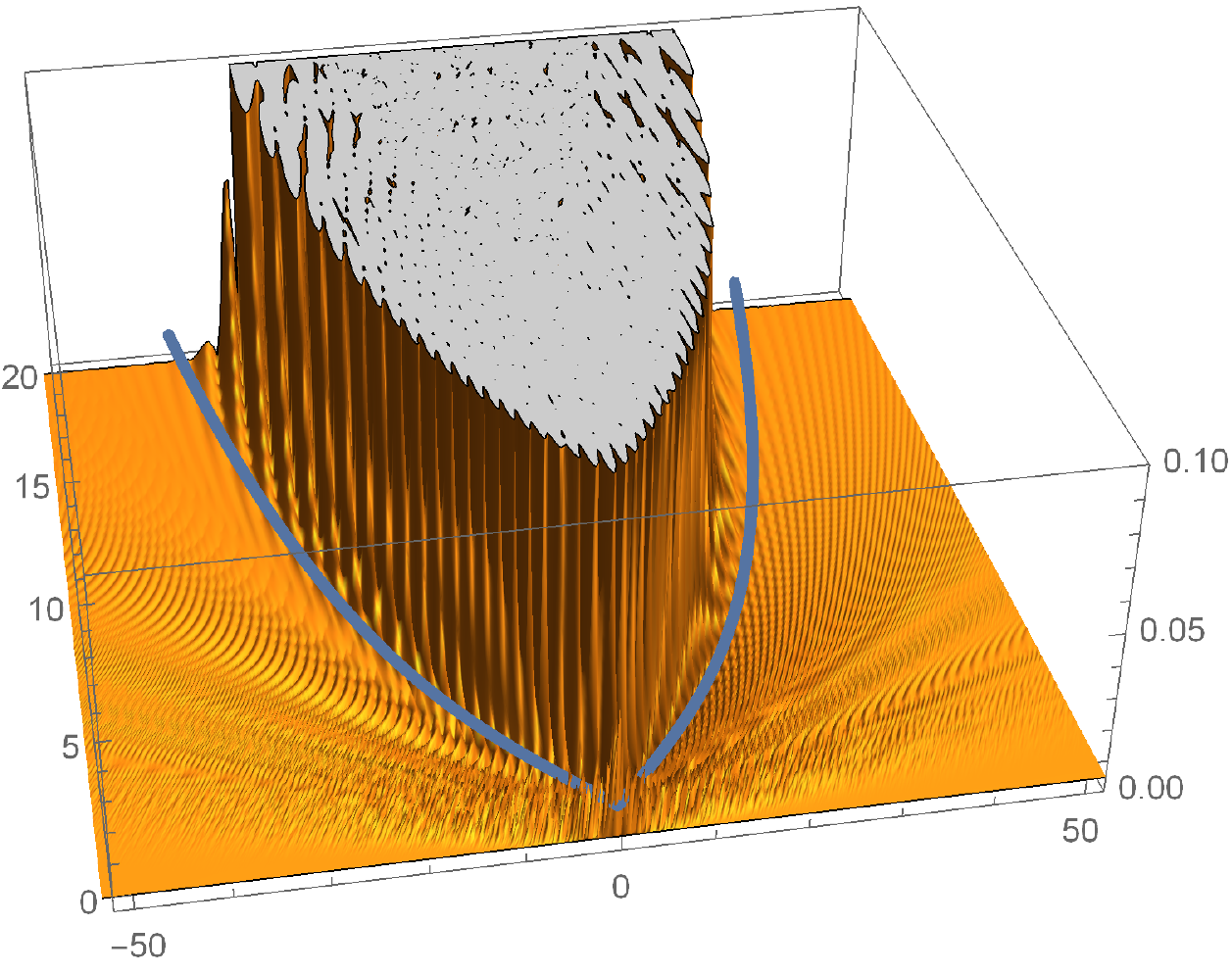}}
\put(35,80){(a)}
\put(155,80){(b)}
\end{picture}
\caption{(a) Plot of the numerical solution of the NLS equation (\ref{eq1}) for the
initial condition (\ref{eq25b}) with $a=1.5$, $l=20$, and a tiny perturbation
of density at $x=0$. (b) Plot of $|\rho(x,t)-\rho_b(x,t)|$ with subtracted background
evolution (\ref{eq29b}), (\ref{eq30b}) from the exact solution $\rho(x,t)$.
Analytical paths of the instability fronts are
shown by blue lines.
 }
\label{fig2}
\end{center}
\end{figure}

The formulas (\ref{eq29b}), (\ref{eq30b}) show that it is convenient to choose $\xi$
as a parameter along the front's path. The group velocity of linear waves
propagating in the neighborhood of the point $x(\xi,t)$ with the carrier wave
number $k_m=\sqrt{6\rho}$ at the moment $t$ can also be expressed as a function
of $\xi$ and $t$:
\begin{equation}\label{eq31}
  v_g(\xi,t)=2a\xi\sqrt{1-\frac{l}{at}\frac{\xi}{1+\xi^2}}-
  2\sqrt{\frac{2al\xi}t}.
\end{equation}
If we denote by $x=\ox(\xi(t),t)$ the front's,
then we have $\ox_{\xi}(d\xi/dt)+\ox_t=v_g(\xi,t)$ (see Eq.~(\ref{eq12b}) where
the parameter $\rho$ is replaced by $\xi$), and with account of Eq.~(\ref{eq30b}) this
equation reduces to the differential equation
\begin{equation}\label{eq32}
\begin{split}
  \frac{d\xi}{dt}=&\frac{\xi(1+\xi^2)}  {t[4a\xi t(1+\xi^2)^2+l(1-6\xi^2-3\xi^4)]}
  \Bigg[l(1-\xi^2)\\
  &-4(1+\xi^2)\sqrt{2al\xi t\left(1-\frac{l}{at}\frac{\xi}{1+\xi^2}\right)}\Bigg]
  \end{split}
\end{equation}
for dependence of the parameter $\xi$ along the path. This equation must be solved
with the initial condition $\xi=1$ at $t=0$
which means instant formation of the oscillatory region due to the disturbance.
This equation can be easily solved numerically and substitution of the
solution into Eq.~(\ref{eq30b}) gives us the path of the edge (the interval
$1>\xi>0$ corresponds to the left edge of the oscillatory region). An example of the
resulting plots is shown in Fig.~2. Again we see quite good agreement of our
analytical theory with numerical solution.

In the same way one can consider the problem of propagation of instability fronts
after the focusing moment $t=t_b$. However, the numerical value of this moment
differs from the theoretical one $t_b=l/(2a)$, so one needs to fit this value 
in the initial condition $\xi=1$ at $t=t_b$. After such a fitting we find again 
good enough agreement with the numerical results.

{\it 5. Conclusion.} In this Letter, we have developed the analytical approach to
the problem of propagation of instability fronts in modulationally unstable systems.
The theory is based on the conjecture that such a front propagates with minimal
group velocity of linear waves propagating in such systems. The theory is applied
to the systems whose evolution is described by the focusing NLS equation and
confirmed by comparison with the results of numerical solutions of this equation.

{\it Acknowledgment.}
	The reported study was partially funded by RFBR, project number 20-01-00063.

%

\clearpage
\widetext
\begin{center}
	\textbf{\large Supplemental Materials: Propagation of instability fronts in modulationally unstable systems}
\end{center}
\setcounter{equation}{0}
\setcounter{figure}{0}
\setcounter{page}{1}
\makeatletter
\renewcommand{\theequation}{S\arabic{equation}}



Here we reproduce solutions of Talanov and Akhmanov-Sukhorukov-Khokhlov
in convenient for us form.

{\it 1. Talanov solution.} We substitute the {\it ansatz}
\begin{equation}\label{s1b}
  \rho_b(x,t)=\frac{a^2}{f(t)}\left(1-\frac{x^2}{l^2f^2(t)}\right),\qquad
  u_b(x,z)=\phi(t)x,
\end{equation}
into Eqs.~(\ref{70-2g}) and after equating the coefficients of the like powers
of $x$ we get the system
\begin{equation}\label{s2b}
  \phi(t)=\frac{f_t}{f},\qquad f_{tt}=-\frac{a^2}{l^2}\cdot\frac2{f^2},
\end{equation}
where $f(0)=1,\phi(0)=0$. The second equation should be solved with the initial
condition $f_t=0$ at $f=1$, so we get Eq.~(\ref{eq22}). This equation can be
readily solved with the initial condition $f(0)=1$ to give the solution
(\ref{eqc.150}).

{\it 2. Akhmanov-Sukhorukov-Khokhlov solution.} As was mentioned in the main text,
Eqs.~(\ref{70-2g}) can be solved by the methods of the compressible fluid
dynamics, so we introduce complex Riemann invariants
\begin{equation}\label{eq16}
  r_+=\frac{u}2+i\sqrt{\rho},\qquad r_-=\frac{u}2-i\sqrt{\rho}
\end{equation}
and transform the system (\ref{70-2g}) to a diagonal form
\begin{equation}\label{eq17}
  \frac{\prt r_+}{\prt t}+\frac12(3r_++r_-)\frac{\prt r_+}{\prt x}=0,\qquad
   \frac{\prt r_-}{\prt t}+\frac12(r_++3r_-)\frac{\prt r_-}{\prt x}=0.
\end{equation}
Now we make the hodograph transform, that is consider $t=t(r_+,r_-),x=x(r_+,r_-)$
as functions of the Riemann invariants, so Eqs.~(\ref{eq17}) are reduced to
\begin{equation}\label{eq18}
  \frac{\prt x}{\prt r_+}-\frac12(r_++3r_-)\frac{\prt t}{\prt r_+}=0,\qquad
  \frac{\prt x}{\prt r_-}-\frac12(3r_++r_-)\frac{\prt t}{\prt r_-}=0.
\end{equation}
Following the generalized hodograph method of Ref.~\cite{Tsarev-1991},
we look for the solution of this linear system in the form
\begin{equation}\label{eq19b}
   x-\frac12(3r_++r_-)t=\frac{\prt W}{\prt r_+},\qquad
   x-\frac12(r_++3r_-)t=\frac{\prt W}{\prt r_-},
\end{equation}
and after some manipulations (see, e.g., \cite{kamch-21c}) obtain the Euler-Poisson equation
\begin{equation}\label{eq20b}
  \frac{\prt^2W}{\prt r_+\prt r_-}-\frac1{2(r_+-r_-)}
  \left(\frac{\prt W}{\prt r_+}-\frac{\prt W}{\prt r_+}\right)=0
\end{equation}
for the function $W=W(r_+,r_-)$. This equation should be solved with the
boundary conditions
\begin{equation}\label{eq21}
  \left(\frac{\prt W}{\prt r_+}+\frac{\prt W}{\prt r_+}\right)_{r_-=-r_+}=2x(-ir_+),\qquad
  \left(\frac{\prt W}{\prt r_+}-\frac{\prt W}{\prt r_+}\right)_{r_-=-r_+}=0,
\end{equation}
where $x(-ir_+)$ is obtained by inverting the initial distribution (\ref{eq25b}):
$x=x_0(\sqrt{\rho_0})=x_0(-ir_+)$. Passing to the real variables $r_+=p+iq,r_-=p-iq$,
$q=\sqrt{\rho}$, $p=u/2$, we readily transform Eq.~(\ref{eq20b}) to the form
\begin{equation}\label{eq22b}
  \frac{\prt^2W}{\prt p^2}+\frac{\prt^2W}{\prt q^2}+\frac1q\frac{\prt W}{\prt q}=0
\end{equation}
and the boundary conditions (\ref{eq21}) to the form
\begin{equation}\label{eq23}
  \left.\frac{\prt W}{\prt p}\right|_{p=0}=2x_0(q),\qquad
   \left.\frac{\prt W}{\prt q}\right|_{p=0}=0.
\end{equation}
If the function $W(p,q)$ is found, then Eqs.~(\ref{eq19b}) yield the solution in
implicit form
\begin{equation}\label{eq23b}
  t=\frac1{2q}\frac{\prt W}{\prt p},\qquad x=2pt+\frac1{2}\frac{\prt W}{\prt q}.
\end{equation}

Equation (\ref{eq22b}) can be considered as the Laplace equation written in cylindrical
coordinates for the electrostatic potential $W$ created by the `charge distribution'
\begin{equation}\label{eq24}
  2x_0(q)=l\cdot\ln[(a-\sqrt{a^2-q^2})/q]
\end{equation}
on the disk $q\leq a, p=0$, $q$ and $p$ being the radial and axial coordinates,
correspondingly. Electrostatic problems of that kind are conveniently solved with the use
of spheroidal coordinates $\xi,\eta$ defined by the formulas (see, e.g., \cite{ll-8})
\begin{equation}\label{eq25}
  p=a\xi\eta,\quad q^2=a^2(1+\xi^2)(1-\eta^2),
\end{equation}
where $0\leq\xi$, $-1\leq\eta\leq1$, so that the ellipsoid $q^2/[a^2(1+\xi^2)]+p^2/(a^2\xi^2)=1$
tends to the disk as $\xi\to0$. The Laplace equation (\ref{eq22b}) takes the form
\begin{equation}\label{eq26b}
  \frac{\prt}{\prt\xi}\left[(1+\xi^2)\frac{\prt W}{\prt\xi}\right]+
  \frac{\prt}{\prt\eta}\left[(1-\eta^2)\frac{\prt W}{\prt\eta}\right]=0
\end{equation}
and the boundary conditions read
\begin{equation}\label{eq27b}
  \left.\frac{\prt W}{\prt \xi}\right|_{\xi=0}= -al\eta\cdot\ln\frac{1+\eta}{1-\eta},\qquad
   \left.\frac{\prt W}{\prt \eta}\right|_{\xi=0}=0.
\end{equation}
Equation (\ref{eq26b}) is readily solved by standard method of separation of variables,
so we obtain
\begin{equation}\label{eq28b}
  W(\xi,\eta)=-al\left[\xi\eta\ln\frac{1+\eta}{1-\eta}-2\xi+2\arctan\xi\right].
\end{equation}
Equations (\ref{eq23b}) transformed to the ellipsoidal coordinates yield the formulas
\begin{equation}\label{eq29}
   t=\frac{l}a\cdot\frac{\xi}{(1+\xi^2)(1-\eta^2)},\qquad
   x=l\left[\frac{2\xi^2\eta}{(1+\xi^2)(1-\eta^2)}-\frac12\ln\frac{1+\eta}{1-\eta}\right]
\end{equation}
which determine implicitly $\xi$ and $\eta$ as functions of $x$ and $t$. When they are
found, their substitution into
\begin{equation}\label{eq30}
  \rho=q^2=a^2(1+\xi^2)(1-\eta^2),\qquad u=2p=2a\xi\eta
\end{equation}
yields the distributions of the density and flow velocity (\ref{eq26}).
In this form the solution was obtained in Ref.~\cite{gs-70b} and it is equivalent to the
results of Ref.~\cite{ask-66b}.

\end{document}